\documentclass[12pt]{article}
\usepackage[latin1]{inputenc}
\usepackage[T1]{fontenc}
\usepackage{amsmath}
\usepackage{amsfonts}
\usepackage{amssymb}
\usepackage{color}
\usepackage{graphicx}
\title{Current correlation functions from a bosonized theory in $3/2+1$
dimensions}
\author{C.~D.~Fosco$^a$ and F.~A.~Schaposnik$^b$
\\
~
\\
~
\\
{\normalsize $^a\!$\it Centro At\'omico Bariloche and Instituto
Balseiro,}\\
{\normalsize $\!$\it Comisi\'on Nacional de Energ\'\i a At\'omica, 8400
Bariloche, Argentina}\\
{\normalsize $\!$\it }\\
{\normalsize $^b\!$\it Departamento de F\'\i sica, Universidad
Nacional de La Plata}\\ {\normalsize\it Instituto de F\'\i sica La Plata-CONICET}\\
{\normalsize\it C.C. 67, 1900 La Plata,
Argentina}
 }
\begin{document}
\date{\today}
\maketitle
\begin{abstract}
Within the context of a bosonized theory, we evaluate the current-current
correlation functions corresponding to a massive Dirac field in
$2+1$ dimensions, which is constrained to a spatial half-plane.  
The boundary conditions are imposed on the dual theory, and have the form of
of perfect-conductor conditions. We also consider, for the sake of
comparison, the purely fermionic version of the model and its boundary
conditions, in the large-mass limit.

\noindent We apply the result about the dual theory to the evaluation of
induced vacuum currents in the presence of an external field, in a
spatial half-plane. 
\end{abstract}
Bosonization is a useful tool which, in $1+1$ space-time dimensions,
allows for the solution of some non-trivial Quantum Field Theory models
(see~\cite{Stone:1995ys} for a comprehensive review and
useful references).

For a massive Dirac field in $2+1$ dimensions, the situation we are
concerned with here,  the path integral bosonization framework may be used
to derive the exact bosonization rule
for the current. The
(dual) bosonic action, is gauge-invariant and, in the massive case, local, what determines the
form of the possible terms in a mass expansion~\footnote{In the massless case, 
the bosonic action is non-local.}. 
Thus, to the leading order,
it is a Chern-Simons term, while the next-to-leading one corresponds, in
the Abelian or non Abelian cases, to a (local) Maxwell \cite{FS}-\cite{c}
or Yang-Mills term~\cite{b}, respectively.
We note that the need for the CS term has been shown explicitly, even in a
massless theory, as a consequence of an $\eta$ function regularization,
required to have a consistent gauge invariant theory~\cite{SSWW}.

In a previous work~\cite{Fosco:2018pcq}, we have applied the functional
bosonization approach to a system consisting of a massive Dirac field
constrained to a $2+1$ dimensional spacetime manifold ${\mathcal U}$, with
non-trivial conditions on its boundary ${\mathcal M} \equiv \partial
{\mathcal U}$.  Those conditions, when imposed on the dual (bosonized)
version of the theory, amounted to the vanishing, at each point of
${\mathcal M}$, of the normal component of the (bosonized) current.  The
bosonization rules, formulated in terms of an Abelian gauge field $A_\mu$,
were shown to be the same as in the no-boundary case, while the existence
of the boundary manifested itself through the fact that the gauge field
satisfied {\em perfect-conductor\/} conditions on ${\mathcal M}$.
This is one of the benefits of the procedure: the avoidance of the
calculation of a fermionic determinant with non-trivial
boundary conditions. Indeed, they are converted into conditions for the
gauge field, easier to implement.

The exact bosonization of a $1+1$ dimensional model with a boundary, i.e.,
on a half-line, has been implemented in~\cite{Fuentes:1995ym}.  In this
article, following~\cite{Fosco:2018pcq}, we apply the bosonization
approach above to the calculation of current correlation functions,
in a concrete geometry: a massive Dirac field confined to a spatial half-plane (so that,
following~\cite{Fuentes:1995ym}, we dub the associated space-time as `$3/2
+1$ dimensions'). 

We do not dwell here with a massless theory, where there seems to be, in
principle, no natural mass to use in the expansion, and
the low energy terms can be non-local. In spite of this, the program could be implemented also in this case (see~\cite{Moreno2} for a discussion), by using the renormalization mass scale $\mu$ as the expansion parameter. Our study of a bosonized Dirac field in $3/2+1$ dimensions, which takes into account the leading and sub-leading terms in the mass
expansion, encompasses the evaluation of the current-current correlation function, in the context of functional bosonization. We also apply it to the determination of the induced current in the presence of an external gauge field, presenting the general form of the result, as well as more explicit expressions 
for some particular cases.

We consider a massive Dirac field in $2+1$
dimensions which, in its fermionic incarnation, is described by an Euclidean action
${\mathcal S}_f({\bar\psi},\psi)$, given by:
\begin{equation}\label{eq:defsf}
{\mathcal S}_f({\bar \psi},\psi) \;=\; \int_{\mathcal U} d^3x \, {\bar
	\psi} (\not \! \partial + m) \psi \;,
\end{equation}
on a spacetime manifold ${\mathcal U}$ which, in terms of the
coordinates \mbox{$x = (x_0,x_1,x_2)$}, corresponds to the
space-time region $x_2 > 0$. The current is assumed to vanish along the
normal  direction to the border (see (\ref{eq:bag}) below for a concrete
implementation in the fermionic version).
In the fermionic version of the model, there are many different ways to achieve
the vanishing of the expectation value of the current on the boundary. 
What we shall see, is  that the same dual theory emerges, as soon as one assumes
 that the  boundary conditions on the fermions are such that 
the model inside the region delimited by the boundary is decoupled from the one 
outside.

Dirac's $\gamma$-matrices are Hermitean and, in our conventions, they
satisfy \mbox{$\gamma_\mu  \gamma_\nu = \delta_{\mu\nu} \,+\, i \,
\epsilon_{\mu\nu\lambda} \gamma_\lambda$}.  Letters from the middle of the
Greek alphabet are assumed to run over the values $0,\,1,\,2$. The
Euclidean metric has been assumed to be the identity matrix
$\delta_{\mu\nu}$, and $\epsilon_{\mu\nu\lambda}$ denotes
the Levi-Civita symbol, with $\epsilon_{012} = +1$.   .

The functional bosonization approach, which we briefly review within the
framework of a given geometry,  begins from the conserved Noether
current corresponding to (\ref{eq:defsf}), namely, $J_\mu = {\bar \psi}
\gamma_\mu \psi$, while the existence of the boundary is reflected in the
vanishing of $J_n \equiv \hat{n}_\mu J_\mu\big|_{\mathcal M}$, the normal
component of the current on the boundary ${\mathcal M} \equiv
(x_\parallel,0)$, with $x_\parallel = (x_0,x_1)$, and the (outer) unit
normal $\hat{n}_\mu = - \delta_{\mu 2}$.

To construct the fermionic generating functional, we need to add two
ingredients: first, a term ${\mathcal S}_J$:
\begin{equation}
{\mathcal S}_J(s,J) \;=\; i \, \int d^3x \, s_\mu(x) J_\mu(x) \;,
\end{equation}
which includes a source $s_\mu$, to be able to generate current correlation
functions. The integral above~\footnote{Integrals are assumed to be
unrestricted, unless explicitly stated otherwise. Namely, $\int d^3x
\ldots$ is assumed to be an integral over ${\mathbb R}^{(3)}$, etc.} does
not need to be restricted to ${\mathcal U}$ if one assumes, as we shall do,
that the source $s_\mu$ (a field which is not functionally integrated) vanishes
outside ${\mathcal U}$. 

A second term, ${\mathcal S}_{\mathcal M}$, depending on an auxiliary field
$\xi(x_\parallel)$, is added in order to impose the condition on the normal current:
\begin{equation}
S_{\mathcal M}(\xi, J) \;=\; - i \, \int d^2x_\parallel \, \xi(x_\parallel)
\, J_2(x_\parallel,0) \;,
\end{equation}
which can be also written as a term which couples the fermionic current to
a vector field $c_\mu(\xi,x)$, which is completely determined by the auxiliary
field and the boundary; indeed:
\begin{equation}\label{eq:cap1}
S_{\mathcal M}(\xi, J) \;=\; i \, \int d^3x \, c_\mu(\xi,x) \, J_\mu(x)
\;\;, \;\;\;
c_\mu(\xi, x) \equiv - \delta_{\mu 2} \, \xi(x_\parallel) \,\delta(x_2) \;.
\end{equation}

Note that the functional integral over $\xi$  yields a (functional) $\delta$
of the normal current:
\begin{equation}
\delta_{\mathcal M}[J_n] \;=\; \int {\mathcal D}\xi \; e^{i\,\int
	d^2x_\parallel \, \xi(x_\parallel) \, J_2(x_\parallel,0) }
\;=\; \int {\mathcal D}\xi \; e^{- S_{\mathcal M}(\xi, J)} \;.
\end{equation}
Note that, assuming the constraint above is due to a boundary condition on the Dirac field which completely determines the problem inside ${\mathcal U}$, one can extend the 
fermionic action to the whole of space-time, since the conditions on the
current isolate the problem on ${\mathcal U}$ from the one in its
complement. In that way, a source which has support on ${\mathcal U}$ will
be oblivious to the existence of a fermionic field outside of ${\mathcal
U}$, and the result of the functional integral becomes a product of one depending on the fields inside (and the source) times another one for the fields outside. The latter cancels out when evaluating expectation values.

On the other hand, the important advantage of interpreting $S_{\mathcal M}$ as a
coupling between the current and a field $c_\mu$ stems from the fact that
the fermionic generating functional ${\mathcal Z}(s)$ may be written as
follows:
\begin{equation}\label{eq:defzs}
{\mathcal Z}(s) \;=\; \int {\mathcal D}\psi \,{\mathcal D}{\bar
\psi}\, {\mathcal D}\xi \; e^{-{\mathcal S}_f({\bar \psi},\psi;s+c) } \;,
\end{equation}
with
\begin{equation}
{\mathcal S}_f({\bar \psi},\psi;s)\;=\; \int d^3x \,
{\bar \psi} (\not \! \partial + i \not \!s + m) \psi \;.
\end{equation}
Note that the fermionic fields do not have an explicit dependence on the boundary, in the sense that they are not restricted spatially to the region ${\mathcal U}$.

Following the procedure devised in~\cite{Fosco:2018pcq},  we now
disentangle $s_\mu + c_\mu$ from the fermionic action in (\ref{eq:defzs}).
Note that this step decoupled the Dirac operator from the boundary, and
will allow to evaluate the fermionic determinant in the absence of borders.
Of course, the borders will reemerge in the bosonic theory. 

To that end, we first perform the change of
variables:
\begin{equation}
	\psi(x) \to e^{i \alpha(x)} \psi(x) \;\;, \;\;\;
	{\bar\psi}(x) \to e^{-i \alpha(x)} {\bar\psi}(x) \;,
\end{equation}
and integrate over $\alpha$, to obtain:
\begin{equation}\label{eq:zseq1}
{\mathcal Z}(s) \;=\;  \int {\mathcal D} \alpha \;{\mathcal D}\xi
\;{\mathcal D}\psi \,{\mathcal D}{\bar \psi} \;
e^{-{\mathcal S}_f({\bar \psi},\psi;s+c+\partial\alpha)} \;.
\end{equation}
Then, the integration over $\alpha$ is substituted by one over a vector
field $b_\mu$ $\partial_\mu \alpha \to b_\mu$,
\begin{equation}\label{eq:zseq2}
{\mathcal Z}(s) \;=\;  \int {\mathcal D}b  \; \delta[{\tilde f}_\mu(b)] \,
{\mathcal D}\xi \;{\mathcal D}\psi \,{\mathcal D}{\bar \psi} \;
e^{-{\mathcal S}_f({\bar \psi},\psi;s+c+b)} \;,
\end{equation}
where ${\tilde f}_\mu(b) = \epsilon_{\mu\nu\lambda} \partial_\nu
b_\lambda =0$ ($b_\mu$ is a pure gradient).

The ${\tilde f}_\mu(b) = 0$ condition is implemented by means of the
representation:
\begin{equation}
	\delta[{\tilde f}_\mu(b)] \;=\; \int {\mathcal D}A \;
	e^{\frac{i}{\sqrt{2\pi}}\, \int
d^3x \, A_\mu {\tilde f}_\mu(b)} \;.
\end{equation}
Thus,
\begin{equation}\label{eq:zseq3}
{\mathcal Z}(s) \;=\;  \int {\mathcal D}A \; {\mathcal D}b  \;
	{\mathcal D}\xi \;{\mathcal D}\psi \,{\mathcal D}{\bar \psi} \;
	e^{-{\mathcal S}_f({\bar \psi},\psi;s+c+b) + \frac{i}{\sqrt{2\pi}}
		\,\int d^3x A_\mu {\tilde f}_\mu(b) } \;.
\end{equation}
Finally, we make the shift $b \to b - c - s$, to obtain:
\begin{equation}\label{eq:zseq4}
{\mathcal Z}(s) \;=\;  \int {\mathcal D}A \; {\mathcal D}b  \;
	{\mathcal D}\xi \;
	e^{-W(b) + \frac{i}{\sqrt{2\pi}} \int d^3x A_\mu
	[ {\tilde f}_\mu(b) - {\tilde f}_\mu(c) - {\tilde f}_\mu(s)]} \;,
\end{equation}
where $W(b)$ denotes the effective action:
\begin{equation}\label{eq:defzb}
	e^{-W(b)} \;=\; \det (\not \! \partial + i \not \! b + m) \;,
\end{equation}
which is to be evaluated with trivial boundary conditions, understanding by that that 
the region is the whole $2+1$-dimensional spacetime, with the standard conditions for a vacuum to vacuum Euclidean transition amplitude.

This leads to a bosonized representation for the generating functional,
which may be rendered as follows:
\begin{equation}\label{eq:zs1}
	{\mathcal Z}(s) \;=\;  \int {\mathcal D}A \, {\mathcal D}\xi \;
		e^{-{\mathcal S}_B(A) \,-\, \frac{i}{\sqrt{2\pi}}\int d^3x c_\mu
\epsilon_{\mu\nu\lambda} \partial_\nu A_\lambda
- \frac{i}{\sqrt{2\pi}}\int d^3x s_\mu \epsilon_{\mu\nu\lambda}
\partial_\nu A_\lambda }\;,
\end{equation}
where the bosonized action ${\mathcal S}_B(A)$ is determined by the expression:
\begin{equation}\label{eq:zs2}
e^{- {\mathcal S}_B(A)} \;=\;  \int {\mathcal D}b  \,
e^{-W(b) + \frac{i}{\sqrt{2\pi}} \int d^3x  \, A_\mu {\tilde f}_\mu(b)} \;.
\end{equation}

This leads to the bosonization rule:
\begin{equation}
	J_\mu(x)  \;\to\;  \frac{1}{\sqrt{2\pi}}
	\epsilon_{\mu\nu\lambda}\partial_\nu A_\lambda  \equiv {\mathcal
	J}_\mu(x) \;,
\end{equation}
with a bosonized action ${\mathcal S}_B(A)$ yet to be determined.  Since
that depends on the knowledge of $W(b)$, an exact expression of which is
unknown, we use a possible approximation to it. The usual approach is to
use a large-mass expansion, retaining just the leading contribution,
a Chern-Simons (CS) term. This term is $m$-independent. Since we are
interested here in dealing with a situation where there is another scale
present, namely, the distance to the boundary, and to allow for a possible
interplay, we will also include the next-to-leading term, which has the
form of a Maxwell action:
\begin{equation}\label{eq:wbexp}
W(b) \;=\; \int d^3x \Big[ \pm \frac{i}{4\pi} \,
	\epsilon_{\mu\nu\lambda} b_\mu \partial_\nu b_\lambda
	\,+\, \frac{1}{48 \pi |m|} f_{\mu\nu}(b)  f_{\mu\nu}(b)
\,+\, {\mathcal O}(1/m^3) \Big] \;,
\end{equation}
where the parity-breaking term has a $\pm$ sign, a reflection of the parity anomaly~\cite{gamboa}.

Inserting this into the expression for the bosonized action ${\mathcal
S}_B(A)$, (\ref{eq:zs2}), and working consistently up to the same order in
the mass expansion, leads to:
\begin{equation}\label{eq:sba}
	{\mathcal S}_B(A) \;=\; \int d^3x \Big[ \mp \frac{i}{2} \, A_\mu
		\epsilon_{\mu\nu\lambda} \partial_\nu A_\lambda  \,+\,
	\frac{1}{24|m|} F_{\mu\nu}(A) F_{\mu\nu}(A) \Big] \;.
\end{equation}

Recalling then (\ref{eq:zs1}), the generating functional ${\mathcal Z}(s)$
requires the evaluation of an $A_\mu$ integral including the perfect-conductor
constraint, what is implemented by the auxiliary field. That integral may
be exactly calculated, for example by integrating out $A_\mu$ firstly, and
then over $\xi$ (a Gaussian).

The integral over $A_\mu$, may be put in the form:
\begin{equation}
	\int {\mathcal D}A \;
		e^{-{\mathcal S}_B(A) \,-\, \frac{i}{\sqrt{2\pi}}\int d^3x A_\mu
	\epsilon_{\mu\nu\rho} (\partial_\nu c_\rho - \partial_\nu s_\rho)}\;,
\end{equation}
where $S_B(A)$ is the action (\ref{eq:sba}).  It is convenient to write
formally  this action (using a shorthand notation for the integrals) in the following way
\begin{equation}
	{\mathcal S}_B(A) \;=\; \frac{1}{2} \, \int_{x,x'} A_\mu(x) \, {\mathcal
	K}_{\mu\mu'}(x,x') \, A_{\mu'}(x')  \;,
\end{equation}
with the kernel:
\begin{equation}
	{\mathcal K}_{\mu\mu'}(x,x') \;=\; \pm i 
	\epsilon_{\mu\lambda\mu'} \partial_\lambda^x + \frac{1}{6 |m|}
	(-\partial_x^2 \delta_{\mu\mu'} + \partial_\mu^x \partial_{\mu'}^x) \delta(x-x') \;,  
\end{equation}
where we have explicitly indicated which argument of the $\delta$ function
the derivatives act upon.

Note that the integral is a Gaussian in terms of $A_\mu$, which is coupled
to a vector field which has a vanishing divergence. 
To calculate the integral, it is convenient to decompose the kernel into
orthogonal projectors; that can be done by starting from the fact that it
can be written in terms of the Fourier space tensors:
\begin{equation}
 P_{\mu\nu}(k) \;=\; \delta_{\mu\nu} - \frac{k_\mu k_\nu}{k^2} \;\;,\;\;\;
 Q_{\mu\nu}(k) \;=\; \varepsilon_{\mu\lambda\nu}
	\frac{k_\lambda}{|k|} \;. 
\end{equation}
These tensors satisfy  relations which in a matrix notation, adopt the
form:
\begin{equation}
P^2 \,=\, P \;,\;\;\; Q^2 \,=\, -P \;,\;\;\; P Q \,=\, Q P \,= Q \;.
\end{equation}
They can then be used to build  a complete set of orthogonal projectors for the
space of $3 \times 3$ Hermitian matrices, which naturally arise in the
Fourier representation. Their orthogonality allows one to deal with each
invariant subspace separately, decomposing the original problem a set of
one-dimensional decoupled problems. 

Taking into account the relations above, we see that, defining $P^{\pm}
\equiv \frac{P \pm i Q}{2}$ and  $P' \equiv I -
P$ ($I$ denotes the identity matrix):
\begin{align}
& P^+ + P^- + P' \,=\, I \;,\;\;  (P^{\pm})^2 \,=\,P^{\pm} \;,\;\;P'^2
\,=\,P'\;, \nonumber\\
& P^+ P^- \,=\,P^- P^+ \,=\,P^{\pm} P' \,=\,P' P^{\pm} \,=\, 0 \;.       
\end{align}

Then, using the Fourier representation, we have for the kernel:
\begin{equation}
	{\mathcal K}\;=\; \pm |k| \, ( P^+ - P^-)  + \frac{k^2}{6 |m|} (P^+
	+ P^-)\;,  
\end{equation}
again in a matrix notation. Gauge fixing can be implemented by adding a term
$\frac{\lambda}{2} (\partial \cdot A)^2$ to the bosonized action. This
amounts to adding to ${\mathcal K}$ an extra term:
\begin{equation}
	{\mathcal K} \,\to \,{\mathcal K}'\;=\; {\mathcal K}  \,+\, \lambda \,k^2 \, P' \;.
\end{equation}
Then, the integral over $A_\mu$ yields:
$$
\int {\mathcal D}A  e^{-{\mathcal S}_B(A) \,-\, \frac{i}{\sqrt{2\pi}}\int d^3x A_\mu
\epsilon_{\mu\nu\rho} (\partial_\nu c_\rho - \partial_\nu s_\rho)}
$$
\begin{equation}
\,=\, \exp\big\{ -\frac{1}{4\pi} \int_{x,x'} 
	\epsilon_{\mu\nu\rho} \partial_\nu (c_\rho(x) - s_\rho(x))
	[{\mathcal K}']^{-1}_{\mu\mu'}(x,x') 
	\epsilon_{\mu'\nu'\rho'} \partial'_{\nu'} (c_{\rho'}(x') -
	s_{\rho'}(x')) \big\} \;, 
\end{equation}
where, using the algebraic relations satisfied by the projectors, we see
that, in Fourier space, 
\begin{equation}
	[{\mathcal K}']^{-1}\;=\; (\pm |k| + \frac{k^2}{6 |m|})^{-1} P^+
	\,+\, (\mp |k| + \frac{k^2}{6 |m|})^{-1} P^- 
	\,+\,\frac{1}{\lambda k^2} \, P' \;.
\end{equation}
It may be seen that $P'$ does not contribute, because $\epsilon_{\mu\nu\rho}
(\partial_\nu c_\rho - \partial_\nu s_\rho)$ has zero divergence.
Indeed, the result is independent of any
gauge fixing, and becomes the exponential of a quadratic action. This
quadratic action will evidently  contain a term with two $c_\mu$ fields, one with two
$s_\mu$ fields, and a term which mixes them both. The $c_\mu$ is dependent
on the boundary (recall (\ref{eq:cap1})).

The term quadratic in $s_\mu$ is independent of the
boundary. There only remains to integrate out $\xi$, which is again a
Gaussian. This produces a term which does depend on the boundary, since
$c_\mu$ does.  

Adding the previously described  contributions, the  result may be presented as follows:
\begin{equation}\label{eq:zsf1}
	{\mathcal Z}(s) \;=\; e^{-T(s)} \;,
\end{equation}
where
\begin{align}
	T(s) &= \frac{1}{2} \int_{x,x'} s_\mu(x) \, \Pi_{\mu\mu'}(x,x')
	\,s_{\mu'}(x') \nonumber\\
	\Pi_{\mu \mu'} &=  \Pi^{(1)}_{\mu\mu'}(x,x') \,+\, \Pi^{(2)}_{\mu\mu'}(x,x') \;,
\end{align}
with $\Pi^{(1)}_{\mu\mu'}$ and  $\Pi^{(2)}_{\mu\mu'}$ denoting qualitatively
different contributions: $\Pi^{(1)}_{\mu\mu'}$ is identical to the
contribution one would obtain for a Dirac field in the absence of
boundaries. $\Pi^{(2)}_{\mu\mu'}$, on the other hand, depends on the existence of the
boundary.  Therefore, it cannot be translation invariant along the
$x_2$ coordinate. We have found it convenient to represent both $\Pi^{(1)}$
and $\Pi^{(2)}$ in terms of their Fourier transforms with respect
to the $x_\parallel$ coordinates (for which there is translation
invariance). 

There is a technical detail here: since there is translation invariance
along just two of the three spacetime coordinates, and parity is broken,
the usual procedure to integrate out Gaussians involving a gauge field
had to be generalized. Indeed, following the approach of decomposing the quadratic
form in the Gaussian integral into terms involving all the possible tensors
compatible with the symmetry, and assuming that indices from the beginning of the Greek alphabet ($\alpha$, $\alpha'$, \ldots) run over the values $0$ and $1$, 
the explicit form of those terms (obtained by Gaussian integration) may be shown to be:
\begin{equation}
	\Pi^{(1,2)}_{\mu\mu'}(x,x') \;=\; \int
	\frac{d^2k_\parallel}{(2\pi)^2} \, e^{i k_\parallel \cdot
	(x_\parallel - x'_\parallel)} \;
\widetilde{\Pi}^{(1,2)}_{\mu\mu'}(k_\parallel; x_2,x'_2)
\end{equation}
with:
\begin{align}
\widetilde{\Pi}^{(1)}_{\mu\mu'}(k_\parallel; x_2,x'_2) \;= &
	\frac{3}{\pi} |m| \; \Big\{ - \delta_\mu^\alpha  \delta_{\mu'}^{\alpha'}
\; \Big[ \delta_{\alpha \alpha'}  \frac{(6 m)^2}{2\sqrt{k_\parallel^2 + (6 m)^2}}
\, + \, \frac{k_\alpha k_{\alpha'}}{2 \sqrt{k_\parallel^2 + (6 m)^2}}
\nonumber\\
&\,\pm\, 3 i \,|m| \, \epsilon_{\alpha \alpha'} \, {\rm sgn}(x_2-x'_2) \Big] \nonumber\\
\, + & \delta_\mu^\alpha \delta_{\mu'}^2
\big[ - \frac{i}{2} {\rm sgn}(x_2-x'_2) \delta_{\alpha\alpha'}
	\pm \frac{3 |m|}{\sqrt{k_\parallel^2 + (6 m)^2}}
\epsilon_{\alpha\alpha'} \big] k_{\alpha'} \nonumber\\
\, + & \delta_\mu^2 \delta_{\mu'}^\alpha
\big[ - \frac{i}{2} {\rm sgn}(x_2-x'_2) \delta_{\alpha\alpha'}
	\mp \frac{3 |m|}{\sqrt{k_\parallel^2 + (6 m)^2}}
\epsilon_{\alpha\alpha'} \big] k_{\alpha'} \nonumber\\
\, + & \delta_\mu^2 \delta_{\mu'}^2 \, \frac{k_\parallel^2}{2\sqrt{k_\parallel^2 + (6 m)^2}}
\Big\} \times e^{- \sqrt{k_\parallel^2 + (6 m)^2} |x_2 - x'_2|} \;,
\end{align}
and
\begin{align}
\widetilde{\Pi}^{(2)}_{\mu\mu'}(k_\parallel; x_2,x'_2) \;= - &
	\frac{3}{\pi} |m| \; \Big\{ \delta_\mu^\alpha  \delta_{\mu'}^{\alpha'}
		\Big[ - \frac{(6 |m|)^2}{2\sqrt{k_\parallel^2 + (6 m)^2}} \delta_{\alpha \alpha'}
\,+ \, \frac{k_\alpha k_{\alpha'}}{2 \sqrt{k_\parallel^2 + (6 m)^2}}
\nonumber\\
\,+ & \frac{(6 m)^2}{\sqrt{k_\parallel^2 + (6 m)^2}}
\,\frac{k_\alpha k_{\alpha'}}{k_\parallel^2}
\pm \frac{3 |m| i}{k_\parallel^2} \, \big(k_\alpha \epsilon_{\alpha'
\beta'} k_{\beta'}  + k_{\alpha'} \epsilon_{\alpha \beta} k_{\beta}\big) \Big]
\nonumber\\
\, + & \delta_\mu^\alpha \delta_{\mu'}^2
\big[ - \frac{i}{2}  \delta_{\alpha\alpha'}
	\pm \frac{3 |m|}{\sqrt{k_\parallel^2 + (6 m)^2}}
\epsilon_{\alpha\alpha'} \big] k_{\alpha'} \nonumber\\
\, - & \delta_\mu^2 \delta_{\mu'}^\alpha
\big[ - \frac{i}{2} \delta_{\alpha\alpha'}
	\pm \frac{3 |m|}{\sqrt{k_\parallel^2 + (6 m)^2}}
\epsilon_{\alpha\alpha'} \big] k_{\alpha'} \nonumber\\
\, + & \delta_\mu^2 \delta_{\mu'}^2 \, \frac{k_\parallel^2}{2\sqrt{k_\parallel^2 + (6 m)^2}}
\Big\} \times e^{- \sqrt{k_\parallel^2 + (6 m)^2} |x_2 + x'_2| } \;,
\end{align}
where we assumed that $x_2 > 0$ and $x'_2 > 0$ (which corresponds to the region
of interest). The explicit form of each contributions has been obtained by a lengthy but otherwise straightforward procedure, namely, by taking into account the form of the orthogonal projectors arising in the inverted kernel, and decomposing them in order to take into account the reduced symmetry in the system due to the dependence on the boundary. In particular, the integrals over the second component of the momentum have been performed, using residues, in order to express the result in a mixed Fourier representation.

We have explicitly checked that each term,
$\widetilde{\Pi}^{(1)}_{\mu\mu'}$ and $\widetilde{\Pi}^{(2)}_{\mu\mu'}$,
satisfies a Ward identity separately. Namely,
\begin{equation}
\left\{
	\begin{array}{ccc}
i k_\alpha \widetilde{\Pi}^{(1,2)}_{\alpha\alpha'}(k_\parallel; x_2,x'_2)
\,+\, \partial_{x_2}\widetilde{\Pi}^{(1,2)}_{2 \alpha'}(k_\parallel;
x_2,x'_2) &=&\;0  \,,\\
i k_\alpha \widetilde{\Pi}^{(1,2)}_{\alpha 2}(k_\parallel; x_2,x'_2)
\,+\, \partial_{x_2}\widetilde{\Pi}^{(1,2)}_{2 2}(k_\parallel;
x_2,x'_2) &=&\;0  \;.
\end{array}
\right.
\end{equation}
Note that the full vacuum polarization, the sum of both terms, should satisfy that constraint, since the current is topologically conserved, and the vanishing of the normal current is compatible with conservation; indeed, it follows from current conservation and the divergence theorem. The fact that {\em each contribution satisfies the identity} can be deduced from the  property that one of them satisfies that identity by itself, since it is identical to the one for a conserved current in the absence of boundaries.

In a mass expansion, and keeping just the leading and sub-leading terms,
one sees that those two objects adopt the form:
\begin{align}
\widetilde{\Pi}^{(1)}_{\mu\mu'}(k_\parallel; x_2,x'_2) \;=\;
\frac{1}{2\pi} & \Big\{ \delta_\mu^\alpha  \delta_{\mu'}^{\alpha'}
\;
\Big[
\mp i  \epsilon_{\alpha \alpha'} \partial_{x_2} \,+\,\frac{1}{6|m|} \big(
	 k_\parallel^2 \delta_{\alpha \alpha'} -
	k_\alpha k_{\alpha'}  - \partial_{x_2}^2  \delta_{\alpha \alpha'}\big)
\Big]
\nonumber\\
+ \delta_\mu^\alpha  \delta_{\mu'}^2 \;
\big(
	\mp  \epsilon_{\alpha \alpha'} k_{\alpha'}
	\,+ & \frac{i k_\alpha}{6|m|}  \partial_{x_2}
\big) \,+\, \delta_\mu^2  \delta_{\mu'}^{\alpha'}
\;
\big(
	\pm  \epsilon_{\alpha' \alpha''} k_{\alpha''}
	\,+\,\frac{i k_{\alpha'}}{6|m|}  \partial_{x_2}
\big) \nonumber\\
	+ \, \delta_\mu^2 \delta_{\mu'}^2 \, \frac{k_\parallel^2}{6 |m|} &
	\Big\}  \delta_m(x_2-x'_2) \;,
\end{align}
and
\begin{align}
\widetilde{\Pi}^{(2)}_{\mu\mu'}(k_\parallel; x_2,x'_2) =
	& - \frac{1}{2\pi} \; \Big\{ \delta_\mu^\alpha  \delta_{\mu'}^{\alpha'} \;
\big[ \delta_{\alpha \alpha'} - 2 \frac{k_\alpha k_{\alpha'}}{k_\parallel^2}
\mp \frac{i}{k_\parallel^2} (k_\alpha \epsilon_{\alpha' \alpha''}
k_{\alpha''} +  k_{\alpha'}  \epsilon_{\alpha \alpha''}
	k_{\alpha''})\big]\partial_{x_2} \nonumber\\
&+\, \delta_\mu^2  \delta_{\mu'}^{\alpha'}
 \big[
(i \delta_{\alpha' \alpha''} \mp \epsilon_{\alpha' \alpha''} )
	k_{\alpha''} \big]
 \,+\, \delta_\mu^\alpha  \delta_{\mu'}^2
 \big[
(- i \delta_{\alpha' \alpha''} \pm \epsilon_{\alpha' \alpha''} )
	k_{\alpha''} \big] \nonumber\\
& + \, \delta_\mu^2 \delta_{\mu'}^2 \, \frac{k_\parallel^2}{6 |m|}
\Big\}  \delta_m(x_2+x'_2) \;,
\end{align}
where, in the above two expansions, we have introduced
\begin{equation}
	\delta_m(x_2) \;\equiv\; 3 m  e^{- 6 m |x_2|} \;,
\end{equation}
which approximates  Dirac's $\delta$-function in the large-$m$ limit. We
have kept a number of terms which is consistent with the Ward identities
(note that, to verify
this, one must use the property that $- 6 m \delta_m(x_2)$ is an
approximates of $\delta'$.)

Let us  apply the above result to the determination of the induced vacuum
currents in the presence of a border and of an external electromagnetic
field.

We begin by pointing out that  $\Pi_{\mu\mu'}$ satisfies:
\begin{equation}\label{eq:pmumu}
	\Pi_{2 \mu'}(x_\parallel, 0^+ ;x'_\parallel,x'_2) \;=\;0 \;\;,
	\;\;\;\; \Pi_{\mu 2}(x_\parallel, x_2;x'_\parallel,0^+) \;=\;0
	\;\;.
\end{equation}
This is consistent with the boundary conditions imposed on the normal
component of the current. Indeed, the vacuum expectation value of the
current in the presence of an external gauge field $a_\mu$, is given by:
\begin{equation}\label{eq:vevj}
\langle {\mathcal J}_\mu(x) \rangle|_a \;=\;  \frac{\int {\mathcal D}A
\delta_{\mathcal M}[{\mathcal J}_n] \; {\mathcal J}_\mu(x) \;
e^{-{\mathcal S}_B(A)-i \int d^3x a_\mu {\mathcal J}_\mu}}{\int {\mathcal
D}A \delta_{\mathcal M}[{\mathcal J}_n] \, e^{-{\mathcal S}_B(A)}} \;,
\end{equation}
or,
\begin{equation}\label{eq:vevj1}
\langle {\mathcal J}_\mu(x) \rangle|_a \;=\;  - i \int d^3y \,
\Pi_{\mu\nu}(x,y) \, a_\nu(y)  \;.
\end{equation}
Thus, (\ref{eq:pmumu}) guarantees that the expectation value of the normal
component of the current vanishes
on ${\mathcal M}$.
An important point we would like to stress is that, in the presence of
borders, the large mass expansion can be problematic, in the sense that
the boundary conditions involve a limit, and the current correlation
functions contain singular functions. Thus, we argue that in the presence
of boundaries it is safer to take the large-mass limit only after
calculating observables (for example, an induced current).

Let us apply the general result to the evaluation of the $0$-component of
the current, i.e., the charge density, in the presence of a point-like
static magnetic vortex, located at $(x_1,x_2) = (h_1,h_2)$, which is minimally
coupled to the current. Namely, an external field $a_\mu$ such that:
\begin{equation}
	\partial_1 a_2(x) - \partial_2 a_1(x) \,=\, \phi \, \delta(x_1- h_1)
	\delta(x_2 - h_2)  \;,
\end{equation}
where $\phi$ denotes the magnetic flux piercing the plane at the vortex
location.

We chose the gauge: $a_0=0$, $a_1=0$, and $a_2 = \phi \, \theta(x_1 - h_1)
\,\delta(x_2 -h_2)$ ($\theta \equiv$ Heaviside's step function) to find that
$\langle {\mathcal J}_1 \rangle = \langle {\mathcal J}_2 \rangle = 0$, and
\begin{align}\label{eq:vevj0}
	\langle {\mathcal J}_0(x) \rangle|_a &=\;  - i \phi \,
	\int_{-\infty}^{+\infty} dy_0 \int_h^{+\infty} dy_1 \;
	\Pi_{02}(x_0,x_1, x_2; y_0, y_1, h_2) \nonumber\\
	 &=\;  - \phi \,
	 \int_{-\infty}^{+\infty} \frac{dk_1}{2\pi} \; \frac{e^{i k_1 (x_1
	 -h_1)}}{k_1 - i \epsilon} \,
	 \Big[ \widetilde{\Pi}^{(1)}_{02}(0,k_1;x_2,h_2) \,+\,
	 \widetilde{\Pi}^{(2)}_{02}(0,k_1;x_2,h_2) \Big]\;.
\end{align}

Using the explicit form of $\widetilde{\Pi}^{(1,2)}$, we see that:
\begin{align}\label{eq:vevj01}
	\langle {\mathcal J}_0(x) \rangle|_a &=\;  \mp \phi \, \frac{(3
	m)^2}{\pi} \; \int_{-\infty}^{+\infty}
	\frac{dk_1}{2\pi} \; \frac{e^{i k_1 (x_1 -h_1)}}{\sqrt{k_1^2 + (6
	m)^2}} \; \nonumber\\
	& \times \,\Big( e^{ -\sqrt{k_1^2 + (6 m)^2}|x_2-h_2|} \,-\,
	 e^{ -\sqrt{k_1^2 + (6 m)^2}|x_2+h_2|} \Big)\;.
\end{align}
From this, we see that the interplay between boundary conditions and parity
breaking implies that the induced charge density vanishes at the boundary
$x_2=0$, since it is the sum of two contributions, one of them being the `image'
(in an electrostatic analogy) of the other.

We see that the infinite-mass limit yields the result,
\begin{equation}\label{eq:vevj02}
	\langle {\mathcal J}_0(x) \rangle|_a \;=\; \to  \mp
	\frac{\phi}{2\pi}  \, \delta(x_1 -h_1) \;
	\big[
		\delta(x_2 - h_2) - \delta(x_2 + h_2)
	\big] \;,
\end{equation}
which can be understood as containing the sum of two contributions: one that is the usual 
charge density induced by a flux, when there is a Chern-Simons term, and the other is due 
to an ({\em image}) contribution, in the electrostatic sense, and due to the presence of the conducting plane.

Let us also consider the induced vacuum current  in the presence of a
electric field of magnitude $E$ in the direction of the $x_2$ coordinate.
Using the gauge field choice $a_0(x_2) = - E \, x_2$, it is
straightforward to show that the only non-vanishing component of the
current is along the $x_1$ direction: a parity-breaking effect.
Since the gauge field is static and translation-invariant along
$x_1$, one sees that:
\begin{equation}
\langle {\mathcal J}_1(x_2) \rangle|_a \;=\;- i \, E \,
	\int_0^{\infty}dx'_2 \,x'_2\,\widetilde{\Pi}_{10}(0; x_2, x'_2)
	\;.
\end{equation}
Inserting the form of $\widetilde{\Pi}_{10}(0; x_2, x'_2)$, we see that:
\begin{equation}
\langle {\mathcal J}_1(x_2) \rangle|_a \;=\;  \mp \,\frac{3 m^2}{\pi}
	\; E \, \delta_m(x_2) \;,
\end{equation}
i.e., a Hall current exponentially concentrated on the border.

\section*{Discussion}\label{sec:concl}
A first issue that we comment here is the form of the current-current correlation
function, from the point of view of the fermionic theory. The contribution
of a massive fermion may be written in terms of the fermion propagator
$\widetilde{S_F}(k_\parallel;x_2,x'_2)$,
which satisfies bag-like boundary conditions. For the case at hand,
that condition adopts the form:
\begin{equation}\label{eq:bag}
	(1 + \gamma_2) \widetilde{S_F}(k_\parallel;0^+,x'_2) \;=\; 0 \;.
\end{equation}
Therefore,
\begin{equation}
\widetilde{\Pi}_{\mu\mu'}(k_\parallel; x_2,x'_2) \;=\; -
\int \frac{d^2p_\parallel}{(2\pi)^2} \,
{\rm tr} \Big[ \gamma_\mu \widetilde{S_F}(p_\parallel + k_\parallel;x_2,x'_2)
\gamma_{\mu'} \widetilde{S_F}(p_\parallel ;x'_2,x_2) \Big] \;.
\end{equation}
Following the massive version of the procedure followed
in~\cite{Fosco:2004cn} for the massless case, it is rather straightforward
to show that the fermion propagator is given by:
\begin{equation}
\widetilde{S_F}(p_\parallel + k_\parallel;x_2,x'_2)\;=\;
\widetilde{S_F}^0(p_\parallel + k_\parallel;x_2,x'_2)\,+\,
\widetilde{S_F}^1(p_\parallel + k_\parallel;x_2,x'_2)\;,
\end{equation}
where
\begin{align}
\widetilde{S_F}^0(p_\parallel ;x_2,x'_2) &=\;
\frac{1}{2} \big[ \gamma_2 \, {\rm sign}(x_2-x'_2) \,+\,
U(p_\parallel) \big] \; e^{- \omega(p_\parallel) |x_2 - x'_2|}
\nonumber\\
\widetilde{S_F}^1(p_\parallel ;x_2,x'_2) &=\,
	\frac{1}{2} \; \frac{ 1 - U(p_\parallel)}{ 1 + U(p_\parallel)} \,
	\big[ U(p_\parallel) -  \gamma_2 \big] \,
	U(p_\parallel)  \; e^{- \omega(p_\parallel) |x_2 + x'_2|} \;,
\end{align}
where $\omega(p_\parallel) = \sqrt{p_\parallel^2 + m^2}$, and
$U(p_\parallel) = (- i \not \!\! p_\parallel + m)/\omega(p_\parallel)$.
Besides the standard, perturbative contribution of a massive fermion, one
should include the parity-anomaly term. The form of the anomalous
contribution, on the other hand,  is again a
local Chern-Simons term. Indeed, it may only proceed from the UV-divergent
part of the calculation. And that corresponds to a fermion loop involving
just the $\widetilde{S_F}^0$ term, in the large-mass limit. Indeed, UV
divergences appear in the coincidence ($x_2 \to x'_2$) limit, and
$\widetilde{S_F}^1$ has large-momentum (exponential) suppression for any
$x_2, x'_2 >0$. At the border, it may indeed contribute with a localized
contribution, which is the form of the terms we have found in the bosonized
form of the problem: indeed, $\widetilde{\Pi}^{(2)}$ is non vanishing only
when $x_2=x'_2=0$ (in the large-mass limit).

We have found an expansion for the current-current correlation function
which involves continuous approximations to the $\delta$-function. This
exhibits the role of the  next to leading term included in the
expansion, which here regulates the behaviour of the kernels in the
effective action. Besides, note that the effective action for the boundary modes will be modified, by the inclusion of a width (set by $1/m$). In a mass expansion they will of course correspond to higher derivative terms in the Floreanini-Jackiw action, inherited from the extra terms on the induced action. The large mass limit has been considered in~\cite{Fosco:2018pcq}.

Finally, we have shown that the current-current correlation function may be
expanded, for a large mass, in a way that preserves the Ward identity.
One of the main lessons to be learnt by the present work, reflected in the concrete realization of the Ward identity in a mass expansion, is that the inclusion of the boundary condition {\em after} taking the large mass limit is justified.
Indeed, one might have suspected that the presence of a strong spatial variation at the boundary could have put the procedure in jeopardy. We have shown that not to be the case,
as long as the effective dual theory is considered, and no fermionic operators are introduced in terms of the bosonic field. Should one be able to do that, they should of course reflect 
a stronger dependence on the boundary, in particular on the fermionic boundary condition.
That information is erased in the present treatment.

In recent years, dualities have been applied to analyze different condensed
matter systems, like topological insulators, superconductors, and fractional
quantum  Hall effect systems \cite{SSWW},\cite{w1}-\cite{Witten:2015aoa}.
In these studies, bosonization in $2+1$ dimensions in the presence of a
boundary like the one considered here may be relevant to the applications
\cite{c21},\cite{c22},\cite{c23},\cite{c24}.

\section*{Acknowledgments}
This research was supported by ANPCyT, CONICET,  UNCuyo and UNLP.


\end{document}